\documentclass[11pt]{article}

\usepackage[a4paper, margin=1in]{geometry} 
\usepackage{setspace}
\usepackage{amsfonts}
\usepackage{amssymb}
\usepackage{amsmath}
\usepackage{mathtools}
\usepackage{cite}
\usepackage{graphicx}
\usepackage{booktabs}
\usepackage{float}
\usepackage{multirow}
\usepackage{siunitx} 
\usepackage{xcolor}
\usepackage{algpseudocode}
\usepackage{algorithm}

\date{}

\allowdisplaybreaks

\begin{document}

\title{Variable Rate Lossy Source-Channel Coding over Channels\\ with Feedback}

\author{Timothy~Liu,
        Fady~Alajaji,
        Tam\'as~Linder
\thanks{This work was supported in part by the Natural Sciences and Engineering Research Council (NSERC) of Canada.}
\thanks{The authors are with the Department of Mathematics and Statistics, Queen’s
University, Kingston, ON K7L 3N6, Canada (e-mail: 18tfl1@queensu.ca; fa@queensu.ca; tamas.linder@queensu.ca).}
}

\maketitle

\begin{abstract}
A variable-rate lossy joint source-channel coding scheme for burst-noise communication channels with noiseless feedback is introduced. The scheme comprises a multi-stage channel optimized vector quantization system that dynamically allocates bits via a greedy algorithm among the variable-rate residual quantizers at each stage, based on the channel output sequence received at the encoder through the feedback link, thereby generalizing a prior fixed-rate scheme. Simulations over an $M$-th order Markov noise Polya contagion channel demonstrate that the proposed variable-rate scheme consistently outperforms the fixed-rate scheme, regardless of the channel bit error rate and noise correlation, achieving signal-to-noise ratio gains of up to about 4.5 dB.

\vspace{1em}
\end{abstract}

\section{Introduction}

Shannon pioneered information theory with his separation principle, showing that reliable communication of data sources over noisy channels is possible even when source and channel codes are designed independently, and establishing channel capacity as the fundamental limit on reliable transmission~\cite{6773024}. However, this limit holds only for unlimited blocklengths and delay, making separation impractical in many applications. Joint source-channel coding (JSCC) has therefore emerged as an alternative, achieving a larger error exponent than separate coding, nearly doubling it for both memoryless channels and Markov noise channels with memory~\cite{4385794}. Recent research further highlights JSCC’s importance, showing that recurrent neural network-based models can learn encoder-decoder mappings effective in low-delay analog communication systems~\cite{9954060}.

In practical JSCC systems, feedback from the receiver to the transmitter is often available. This is relevant in scenarios such as a mobile user communicating with a base station, where the downlink can be effectively noiseless (via the use of powerful error correcting codes) while the uplink remains noisy. Although feedback does not increase capacity for discrete memoryless channels~\cite{1056798} or additive-noise channels~\cite{370168}, it enables adaptive JSCC schemes that exploit past outputs, approaching the Shannon limit without additional delay. In~\cite{Kostina2017JSCCFeedback}, the fundamental limits of JSCC with feedback are studied, providing tight finite-blocklength bounds on achievable distortion. The potential benefits due to feedback have motivated practical schemes, such as a low-delay multi-stage lossy JSCC scheme for noiseless feedback channels~\cite{342433} and DeepJSCC-$f$, a deep learning-based approach that iteratively refines image reconstruction using feedback~\cite{9066966}.
Other works have explored JSCC with feedback over fading channels and orthogonal frequency-division multiplexing environments~\cite{Kafedziski1997JointOFDM}, as well as joint source-channel frameworks that incorporate channel-state information feedback via deep learning architectures~\cite{Xu2023DeepJSCCCSI}.

This paper revisits the scheme in~\cite{342433} and proposes an alternative approach. In~\cite{342433}, after each transmission, the decoder’s received sequence is relayed back to the encoder, which uses a quantizer specific to the feedback sequence to refine the initial quantization. At each stage, all quantizers use the same number of bits, but the posterior source density functions conditioned on feedback generally differ in shape and variance, indicating that fixed-rate allocation may be inefficient. We propose a greedy bit allocation algorithm that exploits these differences to distribute bits more efficiently across stages, achieving a signal-to-noise ratio (SNR)  gain of up to 4.5 dB.
\section{Preliminaries}
\subsection{Channel Optimized Vector Quantization}
Channel optimized vector quantization (COVQ) is a robust and well known JSCC scheme introduced in \cite{53739, 1057373}. The goal of COVQ is to map a sequence of $k$-dimensional source vectors $\mathbf{u} \in \mathbb{R}^k$, described by random vector $\mathbf{U}$ with probability density function $f_{\mathbf{U}}$, to a set of predetermined codewords, called centroids, over a noisy channel. In this scheme, perfect knowledge of the channel transition probabilities is assumed. Let $X$ and $Y$ represent the channel input and output indices, respectively, with common alphabet $\mathcal{I} = \{0, 1, \ldots, N-1\}$. Let $\mathcal{S} = \{S_x, x \in \mathcal{I}\}$ and $\mathcal{C} = \{\mathbf{c}_y, y \in \mathcal{I}\}$ be the partition and codebook, respectively. The expected COVQ distortion is
\begin{align}
    D &=  \sum_{x\in \mathcal{I}} \int_{S_x} d'(\mathbf{u}, x)f_{\mathbf{U}}(\mathbf{u})d\mathbf{u},
\end{align}
where
\begin{align}
    d'(\mathbf{u}, x') = \sum_{y \in \mathcal{I}} P(Y = y|X = x')d(\mathbf{u}, \mathbf{c}_y)
\end{align}
and $d(\cdot,\cdot)$ denotes the distortion measure. For this paper, only the squared error distortion, $d(\mathbf{x}, \mathbf{y}) = \lVert \mathbf{x} - \mathbf{y} \rVert^2$ for $\mathbf{x}, \mathbf{y} \in \mathbb{R}^k$, is considered.
An algorithm used to design locally optimal quantizers is the generalized Lloyd algorithm (GLA) \cite{342433} which follows a simple iterative process: fix $\mathcal{C}$ and optimize $\mathcal{P}$ then vice versa until convergence. The algorithm splits optimization into two sub-problems with the following optimality conditions.
\subsubsection{Nearest Neighbor Condition}
The generalized nearest neighbor necessary condition for optimality, from \cite{53739}, states that given a fixed codebook $\mathcal{C}$, an optimal COVQ must have encoding regions that satisfy
\begin{align}
    S_{x} = \{\mathbf{u} \in \mathbb{R}^k: x = \arg\min_{x' \in \mathcal{I}} d'(\mathbf{u}, x')\}, \quad x \in \mathcal{I}
\end{align}
\subsubsection{Centroid Condition}
The generalized centroid condition for optimality, from \cite{53739}, states that given a fixed partition $\mathcal{S}$, an optimal COVQ must have codewords that satisfy
\begin{align}
    \mathbf{c}_y = \arg\min_{\boldsymbol{\omega} \in \mathbb{R}^k} E[d(\mathbf{U}, \boldsymbol{\omega})|Y = y], \quad y \in \mathcal{I}.
\end{align}
Note that the quantizer resulting from the GLA satisfies both necessary conditions for optimality, but is not necessarily a globally optimal quantizer.
\subsection{Adaptive Channel Optimized Vector Quantization}
\label{acovq}
In \cite{342433}, the authors build upon COVQ and introduce a multi-stage JSCC scheme that adapts to channel feedback in a communication system with a noiseless feedback channel. We will refer to this scheme as fixed-rate adaptive channel optimized vector quantization (FR-ACOVQ). Unlike COVQ where quantization and transmission is done in a single step, FR-ACOVQ takes a successive refinement approach over multiple stages, adapting to the channel feedback. In FR-ACOVQ, quantization is split up into $n$ stages and bits are allocated according to vector $\mathbf{b} = (b_1, b_2, \ldots, b_n)$, where $b_i$ represents the number of bits allocated for stage $i$. Let $\mathcal{I}^{(i)} = \{0,1,\ldots, 2^{b_i} -1\}$ be the set of common channel input and output alphabets at stage $i$. Let $X^i = (X_1, X_2, \ldots, X_i)$ and $Y^i = (Y_1, Y_2, \ldots, Y_i)$ be $i$-tuples of random variables representing the channel input and output index sequences, respectively, from stages 1 to $i$.

The first stage of FR-ACOVQ is a $b_1$-bit COVQ (without any modifications) which provides the initial approximation for the source $\mathbf{U} \in \mathbb{R}^k$, with probability density function $f_{\mathbf{U}}$. We denote the codebook and partition for the first stage as $\mathcal{C}^{(1)}$ and $\mathcal{S}^{(1)}$, respectively. After the encoded index is transmitted over the channel, the encoder receives the channel output indices via the noiseless feedback link. At stage $i \geq 2$, additional quantizers are designed for each channel output sequence $y^{i-1} \in \mathcal{I}^{i-1}$, where $\mathcal{I}^{i-1} = \mathcal{I}^{(1)} \times\mathcal{I}^{(2)} \times \cdots \times \mathcal{I}^{(i-1)}.$ These residual quantizers further refine source values that correspond to channel output index sequence $Y^{i-1} = y^{i-1}$ with a density function $f_{\mathbf{U}|Y^{i-1} = y^{i-1}}$. The posterior probability density function can be calculated by applying Bayes theorem as shown in \cite{342433}. We denote the codebook and partition for channel output $Y^{i-1} = y^{i-1}$ as $\mathcal{C}^{(i)}_{y^{i-1}} = \{\mathbf{c}^{(i)}_{y_i|y^{i-1}}, y_i \in \mathcal{I}^{(i)}\}$ and $\mathcal{S}^{(i)}_{y^{i-1}} = \{S^{(i)}_{x_i|y^{i-1}}, x_i \in \mathcal{I}^{(i)}\}$, respectively. The expected distortion of an FR-ACOVQ at stage $i$ consists of the weighted average of all stage-$i$ residual quantizers as follows:
\begin{align}
    D^{(i)} =\sum_{x_i \in \mathcal{I}^{(i)}}\sum_{y^{i-1} \in \mathcal{I}^{i-1}}&\int_{S_{x_i|y^{i-1}}}d'(\mathbf{u}; x_i', \alpha^{i-1}_{y^{i-2}}(\mathbf{u}), y^{i-1}) f_{\mathbf{U}|Y^{i-1}}(\mathbf{u}|y^{i-1}) d\mathbf{u},
\end{align}
where 
\begin{equation}
    \alpha^{i-1}_{y^{i-2}}(\mathbf{u}) = x^{i-1} \in \mathcal{I}^{i-1} \iff \mathbf{u} \in S^{(i-1)}_{x_{i-1}|y^{i-2}} \cap S^{(i-1)}_{x_{i-2}|y^{i-3}}\cap\cdots\cap S^{(1)}_{x_1}, \label{alpha}
\end{equation}    
and 
\begin{align}
    d'(\mathbf{u}; x_i', \alpha^{i-1}_{y^{i-2}}(\mathbf{u}), y^{i-1}) 
    &=  \sum_{y_i \in \mathcal{I}^{(i)}} P\big(Y_i = y_i | X_i = x_i, Y^{i-1} = y^{i-1},
    X^{i-1} = \alpha^{i-1}_{y^{i-2}}(\mathbf{u})\big) d(\mathbf{u}, \mathbf{c}^{(i)}_{y_i|y^{i-1}}).
\end{align}
Similar to COVQ, the residual quantizers in FR-ACOVQ are designed using GLA with the following modified necessary conditions for optimality. In these conditions it is assumed that all partitions and codebooks from the previous $i-1$ stages are fixed.

\subsubsection{Nearest Neighbor Condition}
The nearest neighbor condition, from \cite{342433}, states that for fixed codebook $\mathcal{C}^{(i)}_{y^{i-1}}$ an optimal quantizer has a partition that satisfies
\begin{equation}
     S^{(i)}_{x_i|y^{i-1}} = \bigg\{\mathbf{u} \in \mathbb{R}^k: x_i = \arg\min_{x_i' \in \mathcal{I}^{(i)}} d'(\mathbf{u}; x_i', \\ \alpha^{i-1}_{y^{i-2}}(\mathbf{u}), y^{i-1})\bigg\}, \quad
    x_i \in \mathcal{I}^{(i)}, y^{i-1} \in \mathcal{I}^{i-1}.
\end{equation}
Note that if the channel is memoryless the nearest neighbor condition is equivalent to that of COVQ.

\subsubsection{Centroid Condition}
The centroid condition, from \cite{342433}, states that for a fixed partition $\mathcal{S}^{(i)}_{y^{i-1}}$ an optimal quantizer has a codebook that satisfies
\begin{align}
    \mathbf{c}^{(i)}_{y_i|y^{i-1}} = \arg\min_{\boldsymbol{\omega} \in \mathbb{R}^k} E[d(\mathbf{U}, \boldsymbol{\omega})|Y^i = y^i],\quad 
    y^i \in \mathcal{I}^i.
\end{align}

\subsection{Channel Model}
While COVQ and FR-ACOVQ can be applied to any discrete channel model, in this work we use the finite-memory Polya contagion channel developed in \cite{340476} for all simulations. This channel and its queue-based extension closely approximate other finite-state hidden Markov models (HMMs), such as the Fritchman and Gilbert-Elliott channels~\cite{Zhong2007BinaryQueueChannel}, and real-world fading channels~\cite{4357082}, where burst errors occur rather than independent errors as in memoryless channels. The Polya contagion channel features a binary stationary $M$th-order Markov noise process ${Z_t}_{t=1}^{\infty}$, fully described by three parameters: memory $M \ge 1$, correlation $\delta \ge 0$, and bit error rate $P(Z_t=1)=\epsilon$. As opposed to the aforementioned HMMs, the Polya channel  admits closed-form analytical expressions for both its distribution and its capacity.
The
 joint noise distribution of the Polya channel is given as follows:
\begin{align}
    P(Z^t = (z_1, \ldots, z_t)) = \frac{\prod_{j=0}^{s_t-1}(\epsilon + j\delta)\prod_{j=0}^{t-s_t-1}(1-\epsilon + j\delta)}{\prod_{j=1}^{t-1}(1+j\delta)} \label{polya1}
\end{align}
for $0<t \leq M$, where $s_t = z_1+\ldots+z_t$, $z_l \in \{0,1\}$, $l\ge 1$;
\begin{align}
    P(Z^t = (z_1, \ldots, z_t)) 
    &= L\prod_{j=M+1}^n\left(\frac{\epsilon+{\tilde{s}_{j-1}}}{1+M\delta}\right)^{z_j}\left(\frac{1-\epsilon+(M-\tilde{s}_{j-1})\delta}{1+M\delta} \right)^{1-z_j},
\end{align} 
for $t>M$, where $L = P(Z^{M} = (z_1, \ldots, z_{M}))$ as in~\eqref{polya1} and $\tilde{s}_{j-1} = z_{j-M}+\ldots+z_{j-1}$. Note that when $\delta = 0$, the Polya contagion channel reduces to the memoryless binary symmetric channel with bit error rate~$\epsilon$.

Let $\mathcal{I} = \{0,1,\ldots, 2^b -1\}$ be the common index set for the inputs and outputs of the channel for some positive integer $b$. To obtain the transition probability, $P(Y=y|X=x)$ for $x,y \in \mathcal{I}$, under the Polya contagion channel, we first convert the input $x\in \mathcal{I}$ into a $b$-bit binary vector and transmit it bit-by-bit over the channel, where the noise process $\{Z_l\}_{l=1}^\infty$ is added to it using modulo 2 addition. Specifically, we have that 
   $$\big(\beta_b(y)\big)_l = \big(\beta_b(x)\big)_l \oplus z_l, \qquad
1 \leq l \leq b,$$ where $\beta_\eta:\{0,1,\ldots, 2^\eta-1\} \to \{0,1\}^\eta$ is the function that returns the $\eta$-bit binary vector representation for any integer in $\{0,1,\ldots, 2^\eta-1\}$ and $(\mathbf{v})_l$ is the $l$-th component of binary vector $\mathbf{v} \in \{0,1\}^\eta$. As a result, we have 
\begin{align}
    P(Y= y | X = x) = P\left(Z^b = \beta_b(x) \oplus \beta_b(y)\right),
\end{align}
where $Z^b = (Z_1, Z_2, \ldots, Z_b)$. The transition probabilities in FR-ACOVQ can be obtained similarly. Consider $i \geq 2$ channel input and output sets $\mathcal{I}^{(j)} = \{0,1,\ldots, 2^{b_j} - 1\}$, where $b_j$ is an arbitrary positive integer, for $j = 1,2,\ldots, i$. For some $x_i,y_i \in \mathcal{I}^{(i)}$ and $x^{i-1}, y^{i-1} \in \mathcal{I}^{i-1}$, we have 
\begin{align}
P\big(Y_i = y_i \,\big|\, X_i = x_i,\, &Y^{i-1} = y^{i-1},\, X^{i-1} = x^{i-1}\big) \nonumber \\
&= P\Big( Z^{B_i}_{B_{i-1}} = \beta_{b_i}(x_i) \oplus \beta_{b_i}(y_i) \,\Big|\, \nonumber Z^{B_{i-1}}_{B_{i-2}} = \beta_{b_{i-1}}(x_{i-1})
\oplus\beta_{b_{i-1}}(y_{i-1}), \nonumber \\
&\qquad \qquad \qquad \qquad \ldots, \; Z^{b_1} = \beta_{b_1}(x_1) \oplus \beta_{b_1}(y_1) \Big), \nonumber
\end{align}
where $B_j = \sum_{i = 1}^j b_i$ for $j \geq 1$ and $Z^{j}_l = (Z_l, Z_{l+1}, \ldots, Z_j)$ for $j> l$.

\section{Variable Rate ACOVQ}

\begin{figure}
    \centering
    \includegraphics[width=0.75\linewidth, height=0.4\linewidth]{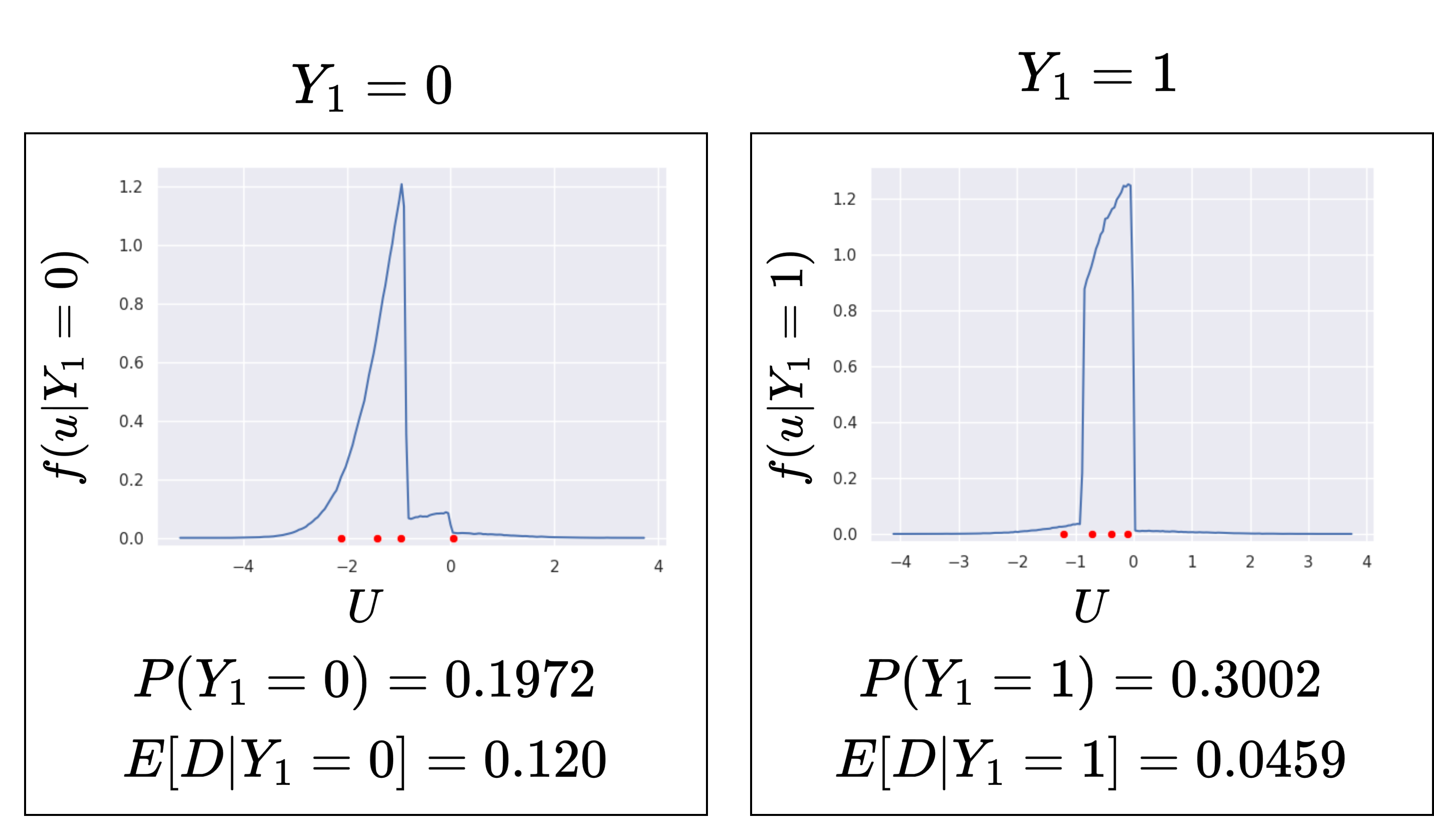}
    \caption{Second stage posterior distributions of FR-ACOVQ with bit allocation $\mathbf{b} = (2,2)$ over the Polya contagion channel with memory $M=1$,  $\epsilon = 0.05$, and $\delta = 5$ (omitting distributions corresponding to $Y_1 = 2$ and $Y_1 = 3$).}
    \vspace{-0.3cm}
    \label{fig:acovq}
\end{figure}


Modern research continues to highlight the benefits of variable-rate coding in practical domains, including audio compression \cite{10889508} and JSCC schemes for image transmission \cite{10743744}. In these areas, dynamically adjusting rates has been shown to improve performance. The same principle can be applied to the FR-ACOVQ scheme. In Figure \ref{fig:acovq}, the posterior distributions and centroids of second-stage quantizers are represented by blue lines and red dots, respectively, for a $\mathbf{b} = (2,2)$ FR-ACOVQ trained on a standard normal distribution. We observe that the distortion from the quantizer corresponding to $Y_1 = 0$ is about three times that of the quantizer corresponding to $Y_1 = 1$. This suggests that better performance can be obtained by allocating fewer bits to quantizers with low variance sources and more bits to ones with high variance sources, while maintaining the same average rate. By allowing bit assignments to vary adaptively across quantizers within a stage, we enable a more flexible and distortion-efficient allocation of resources. In this section, we generalize FR-ACOVQ from a fixed-rate allocation to a variable-rate allocation. 

\subsection{Variable Rate ACOVQ Overview}
We begin with a broad overview of our proposed $n$-stage variable rate ACOVQ (VR-ACOVQ). Similar to FR-ACOVQ, VR-ACOVQ is a multi-stage residual quantization scheme, where the first stage is a basic COVQ. However, after the first stage, bits can be allocated non-uniformly across all quantizers in a given stage. Thus, the codebook sizes for stages after the first can vary based on the previous sequence of channel outputs. As a result, we define a new set $\mathcal{H}^{(i)}$ that specifies the possible values of $X^i$ and $Y^i$ for $i = 1, \ldots, n$, and describe how the VR-ACOVQ is constructed across multiple stages. 

Assume, with a slight abuse of notation, that the first stage is simply a $b$-bit COVQ, for some positive integer $b$, without any modifications. At this stage all source values are quantized with an equal number of bits and $\mathcal{H}^{(1)} = \{0,1,\ldots, 2^{b} -1\}$. Afterwards,
for stages $2 \leq i \leq n$, the construction proceeds recursively in the following 2 steps. These steps occur after receiving the feedback sequences at stage $i-1$ and before training the residual quantizers at stage $i$, using GLA.

\bigskip\noindent
\textbf{Bit allocation:} Given
\begin{align}
    \mathcal{H}^{(i-1)} = \{\mathbf{h}^{(i-1)}_{1}, \mathbf{h}^{(i-1)}_{2}, \ldots, \mathbf{h}^{(i-1)}_{m_{i-1}}\},
\end{align}
assign a bit allocation vector
    \begin{align}
        \boldsymbol{\phi}^{(i)} = (\phi^{(i)}_1, \phi^{(i)}_2, \ldots, \phi^{(i)}_{m_{i-1}}),
    \end{align}
where $m_{i-1} = |\mathcal{H}^{i-1}|$ and $\phi^{(i)}_j$ specifies the number of bits allocated to refine source values corresponding to channel output sequence $Y^{i-1} = \mathbf{h}^{(i-1)}_j$ for $j = 1, \ldots, m_{i-1}$. 

\bigskip\noindent
\textbf{Update channel index sequence set:} given $\mathcal{H}^{(i-1)}$ and $\boldsymbol{\phi}^{(i)}$, construct the new set of channel output index sequences as follows:
\begin{align}
    \mathcal{H}^{(i)} = \bigcup_{j = 1}^{m_{i-1}} \{\mathbf{h}^{(i-1)}_j\} \mathbin\Vert \mathcal{I}_{\phi^{(i)}_j},
\end{align}
where $\mathcal{I}_{N} = \{0,1,\ldots, 2^{N}-1\}$, for some positive integer $N$, and $\mathbin\Vert$ denotes an operator that takes 2 sets of tuples and concatenates all elements from second set to all the elements of the first set (e.g., $\{(0,1)\} \mathbin\Vert \{(2),(3)\} = \{(0,1,2),(0,1,3)\}$).

Let $\bar{\phi}^{(i)} := \sum_{j = 1}^{m_{i-1}}P(Y^{i-1} = \mathbf{h}^{(i-1)}_j)\phi^{(i)}_j$ represent the average rate for stage $i = 1, 2, \ldots, n$. In this paper, we consider $n$-stage VR-ACOVQs constrained by the following bit allocation vector $\boldsymbol{\Phi} = (\Phi^{(1)}, \Phi^{(2)}, \ldots, \Phi^{(n)})$, such that $\bar{\phi}^{(i)} \leq \Phi^{(i)}$ and maximum bit constraint for any quantizer $\Phi_{max}$, such that $\phi^{(i)}_j \leq \Phi_{max}$ for any $i = 1, \ldots, n$ and $j = 1, \ldots, m_{i-1}$.

\subsection{Bit Allocation Algorithm}
We now describe an effective way to determine the number of bits. The FR- and VR-ACOVQ multi-stage, adaptive quantization processes exhibit a tree-like structure, where each path from the root COVQ corresponds to a unique sequence of channel output indices. Finding a globally optimal $\boldsymbol{\phi}^{(i)}$ is a computationally expensive and complex problem, especially in the absence of a closed form distortion-rate function for COVQ at finite rates. Therefore, we adopt an existing greedy bit allocation algorithm for variable rate tree structured quantizers.

In \cite{75264}, Riskin proposed a pruning algorithm for optimizing the bit allocation for variable-rate tree structured quantizers. The algorithm starts with a large tree structured quantizer and greedily removes bits from the ``least efficient" quantizer until only the root quantizer remains. The author notes that the steps in the pruning algorithm can be reversed to create a ``growing" algorithm. In this section, we propose a bit allocation algorithm for VR-ACOVQ based on the ``growing version" of Riskin's pruning algorithm with the addition of a stopping condition to provide an average rate ceiling.

Consider a VR-ACOVQ at stage $i$ and assume we have $m_{i-1}$ unique channel  index sequences received by the encoder via the noiseless feedback link. The bit allocation algorithm takes the following inputs:
\begin{enumerate}
    \item $\mathcal{H}^{(i-1)} = \{\mathbf{h}^{(i-1)}_1, \mathbf{h}^{(i-1)}_2, \ldots, \mathbf{h}^{(i-1)}_{m_{i-1}}\}$, the set of channel output index sequences at stage $i-1$.
    \item Associated probabilities $P(Y^{i-1} = \mathbf{h}^{(i-1)}_j)$ and posterior  density functions $f_{\mathbf{U}|Y^{i-1}}(\mathbf{u}|\mathbf{h}^{(i-1)}_j)$ for $j =1, \ldots, m_{i-1}$.
    \item Average rate constraint $\Phi^{(i)}$ and maximum bit constraint $\Phi_{max}$.
\end{enumerate}
The algorithm outputs $\boldsymbol{\phi}^{(i)} = (\phi^{(i)}_1, \phi^{(i)}_2, \ldots, \phi^{(i)}_{m_{i-1}})$, satisfying $\phi^{(i)}_j \leq \Phi_{max}$ for $j = 1, 2, \ldots, m_{i-1}$ and $\bar{\phi}^{(i)} \leq \Phi^{(i)}$, where $\phi^{(i)}_{j}$ represents the number of bits allocated to quantize source values corresponding to channel output sequence $Y^{i-1} = \mathbf{h}_j$. Let $D(b, \mathbf{h}^{(i)})$ represent the distortion of a $b$-bit residual quantizer associated with the channel output sequence $Y^i = \mathbf{h}^{(i)}$. For brevity, let $p_j := P(Y^{i-1} = \mathbf{h}^{(i-1)}_j)$ for $j = 1, 2, \ldots, m_{i-1}$. The bit allocation algorithm steps are as~follows:
\begin{enumerate}
    \item Set $\phi_{j}^{(i)} = 0$ for all $j = 1, \ldots, m_{i-1}$. This is the initial state of the bit allocation algorithm.  Set $\mathcal{J} = \{1, \ldots, m_{i-1}\}$. The set $\mathcal{J}$ represents the indices of all quantizers whose bit allocation can be incremented without violating any constraints. Elements of $\mathcal{J}$ are removed if a bit increase for the corresponding index violates any constraints.
    \item Set 
    \begin{align*}
        &\lambda_j = \frac{\Delta D^{(i)}}{\Delta \overline{\phi}^{(i)}} \\
        &= \frac{p_j\left(D(\phi^{(i)}_{j}, \mathbf{h}^{(i-1)}_j) - D(\phi^{(i)}_{j} + 1, \mathbf{h}^{(i-1)}_j)\right)}{p_j\left((\phi^{(i)}_{j}+1) - (\phi^{(i)}_{j})\right)} \\
        &= D(\phi^{(i)}_{j}, \mathbf{h}^{(i-1)}_j) - D(\phi^{(i)}_{j} + 1, \mathbf{h}^{(i-1)}_j), \quad \forall j \in \mathcal{J}.
    \end{align*}
    Each element $\lambda_j$ represents the ratio of the decrease in average distortion per increase in average rate by allocating an extra bit to the quantizer corresponding to sequence $\mathbf{h}^{(i-1)}_j$ for all $j = 1, \ldots, m_{i-1}$.
    \item Find $j_{max} = \arg\max_{j \in \mathcal{J}} \lambda_j$. Determine if 
    \begin{align*}
        \sum_{l \in \{1, \ldots, m_{i-1}\} \backslash \{j_{max}\}} p_l \phi^{(i)}_{l} + p_{j_{max}} \left( \phi^{(i)}_{{j_{max}}} + 1 \right) > \Phi^{(i)}
    \end{align*}
    or if ${\phi^{(i)}_{{j_{max}}} = \Phi_{max}}$. The inequality determines whether this increase in allocation violates the average bit allocation constraint for the given stage. If either statement is true, set $\mathcal{J} = \mathcal{J} \backslash j_{max}$ and set $\lambda_{j_{max}} = 0$. Else set $\phi^{(i)}_{{j_{max}}} = \phi^{(i)}_{{j_{max}}} + 1$.
    \item If $\mathcal{J} = \emptyset$ or $\lambda_j \leq 0$ for all $j = 1, \ldots, m_{i-1}$, stop and return $\boldsymbol{\phi}^{(i)}$. Else, repeat steps 2 - 4.
\end{enumerate}%
\subsection{Encoding Complexity Analysis}
The bit allocation algorithm and quantizer training, while expensive, are done offline. Thus, we will focus on the online encoding complexity of VR- and FR-ACOVQ. In \cite{61130}, the authors note that the encoding complexity of COVQ is proportional to the number of nonempty encoding regions. A similar derivation can show the same property holds in ACOVQ, given nonzero codebook sizes. Here we will use the average codebook size as a proxy measure for complexity at stage $i$ of a given VR- and FR-ACOVQ. Let $\mathcal{H}^{(i-1)}, \boldsymbol{\phi}^{(i)}$, $\Phi_{max}$ be the channel index sequence set, bit allocation vector, and maximum bits allocated to a single quantizer for stage $i$ of a VR-ACOVQ, respectively, and let $b_i$ be the channel bit allocation for stage $i$ of a FR-ACOVQ such that $b_i = \bar{\phi}^{(i)}$. Let $\bar{\lVert\mathcal{C}^{(i)}\rVert}_{FR}$ and $\bar{\lVert\mathcal{C}^{(i)}\rVert}_{VR}$ represent the average codebook size for FR- and VR-ACOVQ, respectively. The average codebook size for FR-ACOVQ would simply be $ \bar{\lVert\mathcal{C}^{(i)}\rVert}_{FR} = 2^{b_i}$. Assuming $\phi^{(i)}_j \geq1$ for $j = 1, \ldots, m_{i-1}$ we have
\begin{align}
    \bar{\lVert\mathcal{C}^{(i)}\rVert}_{VR} &= \sum_{j=1}^{m_{i-1}}P(Y^{i-1} = \mathbf{h}^{(i-1)}_j)2^{\phi^{(i)}_j} \\
    & \geq 2^{\sum_{j=1}^{m_{i-1}} P(Y^{i-1} = \mathbf{h}^{(i-1)}_j)\phi^{(i)}_j} \label{jensens}
\end{align}
where \eqref{jensens} holds by Jensen's inequality and the convexity of the function $g(x) = 2^x$. Hence we have that $ \bar{\lVert\mathcal{C}^{(i)}\rVert}_{VR} \geq2^{b_i} \implies \bar{\lVert\mathcal{C}^{(i)}\rVert}_{VR} \geq \bar{\lVert\mathcal{C}^{(i)}\rVert}_{FR}$. However, with a maximum single codebook size $\Phi_{max}$ the average codebook size is upper bounded by $\bar{\lVert\mathcal{C}^{(i)}\rVert}_{VR} \leq \frac{\bar{\phi}^{(i)}}{\Phi_{max}}2^{\Phi_{max}}.$

\section{Simulation Results and Discussion}

We present numerical results for VR-ACOVQ performance as well as the performance gain of VR- over FR-ACOVQ in Tables \ref{tab:tab1} and \ref{tab:tab2}. The VR-ACOVQs were trained with $\boldsymbol{\Phi} = (1,4,1)$ and $\boldsymbol{\Phi} = (4,1,1)$, while the FR-ACOVQs used $\mathbf{b} = (1,4,1)$ and $\mathbf{b} = (4,1,1)$. Both quantizers were trained on at least 4,000,000 samples drawn from a \mbox{1-dimensional} Laplacian distribution with mean zero and scale~1 using the increase-decrease method described in \cite{53739}, with $\Phi_{max} = 8$. Only results from the best performing quantizers were included in the table. The results are given in terms of the source's signal-to-noise ratio (SNR) defined as SNR $= 10 \log_{10} \left(\frac{\sigma^2}{{E[D]}} \right)$ (in dB), where $E[D]$ is the expected distortion of the quantizer and $\sigma^2 =2$ is the variance of the source. In each row, VR and $\Delta$ correspond to the VR-ACOVQ SNR performances (in dB) and the difference between the VR- and FR-ACOVQ SNRs (in dB), respectively.

Results in Tables \ref{tab:tab1} and \ref{tab:tab2} demonstrate that performances for both FR- and VR-ACOVQ schemes improve as the channel memory $(\delta)$ increases for any fixed $\epsilon$, indicating that both schemes effectively exploit channel memory. Additionally, regardless of the channel's $\delta$ and $\epsilon$ values, VR-ACOVQ always outperforms FR-ACOVQ despite never exceeding the average rate of FR-ACOVQ. The gain ranges from $1.361-3.988$ dB for $M=1$ and $1.376-4.527$ for $M=2$. The performance of VR-ACOVQ and its gain over FR-ACOVQ stay relatively consistent from $M=1$ to $M=2$ with some notable exceptions (i.e., $\delta=5,\epsilon=0.05$ and $\delta=10, \epsilon = 0.1$). However, there is no clear overall SNR gain for $M=2$ over $M=1$ and vice versa. Further, variations in performance between different channel memory lengths likely stem from the generalized Lloyd algorithm converging to different local minima, a process highly sensitive to the initial codebooks, and propagating to remaining residual quantizers.

\renewcommand{\arraystretch}{0.5}
\begin{table}
\begin{center}
\caption{VR-ACOVQ SNR performance and gain $(\Delta)$ over FR-ACOVQ for 1-dimensional ($k=1$) memoryless Laplacian source and channel Markov memory $M = 1$.\\}
\label{tab:tab1}

\bigskip

\resizebox{0.7\textwidth}{!}{
\begin{tabular}{ccccccccc}
& & \multicolumn{3}{c}{$\boldsymbol{\Phi} = (1, 4, 1)$} & \multicolumn{3}{c}{$\boldsymbol{\Phi} = (4, 1, 1)$} \\
\cmidrule(r){3-5} \cmidrule(l){6-8}
$\epsilon$   &  & $\delta = 0$ & $\delta = 5$ & $\delta = 10$ &  $\delta = 0$ & $\delta = 5$ & $\delta = 10$  \\
\midrule
\multirow{3}{*}{0.0005} & VR      & 27.064 & 27.834 & 28.635 & 27.423 & 29.221 & 30.031    \\
                        & $\Delta$ &   +2.861 & +2.059 & +2.250 & +3.449 & +3.988 & +3.103 \\
\midrule
\multirow{3}{*}{0.0010} & VR      &  25.831 & 26.756 & 27.612 & 25.727 & 27.898 & 29.301 \\
                        & $\Delta$ & +3.275 & +2.401 & +2.177 & +2.750 & +3.981 & +3.859 \\
\midrule
\multirow{3}{*}{0.0050} & VR      &  22.772 & 23.419 & 24.433 & 22.455 & 23.431 & 25.152  \\
                        & $\Delta$ &   +3.505 & +2.977 & +2.587 & +2.895 & +2.791 & +3.262  \\
\midrule
\multirow{3}{*}{0.0100} & VR      &  20.737 & 22.453 & 23.690 & 21.021 & 21.975 & 23.686   \\
                        & $\Delta$ & +3.305 & +3.772 & +3.413 & +2.750 & +2.884 & +3.441 \\
\midrule
\multirow{3}{*}{0.0500} & VR      & 13.977 & 17.626 & 19.512 & 13.419 & 17.319 & 19.339 \\
                        & $\Delta$ & +2.370 & +3.978 & +3.986 & +1.513 & +3.038 & +3.038  \\
\midrule
\multirow{3}{*}{0.1000} & VR      & 9.409 & 14.867 & 17.069 & 9.957 & 14.750 & 16.651  \\
                        & $\Delta$ &  +1.361 & +3.628 & +3.595 & +1.826 & +2.857 & +2.518 \\
\midrule
\end{tabular}}
\end{center}
\end{table}

\begin{table}[h] 
\begin{center}
\caption{VR-ACOVQ SNR performance and gain $(\Delta)$ over FR-ACOVQ for 1-dimensional ($k=1$) memoryless Laplacian source and channel Markov memory $M = 2$.\\}

\bigskip

\label{tab:tab2}
\resizebox{0.7\textwidth}{!}{
\begin{tabular}{ccccccccc}
& & \multicolumn{3}{c}{$\boldsymbol{\Phi} = (1, 4, 1)$} & \multicolumn{3}{c}{$\boldsymbol{\Phi} = (4, 1, 1)$} \\
\cmidrule(r){3-5} \cmidrule(l){6-8}
$\epsilon$   &  & $\delta = 0$ & $\delta = 5$ & $\delta = 10$ &  $\delta = 0$ & $\delta = 5$ & $\delta = 10$  \\
\midrule
\multirow{3}{*}{0.0005} & VR      & 27.103 & 28.122 & 28.583 & 27.629 & 28.262 & 30.057    \\
                        & $\Delta$ &  +2.756 & +2.399 & +2.143 & +3.538 & +3.568 & +3.276 \\
\midrule
\multirow{3}{*}{0.0010} & VR      & 26.571 & 26.425 & 27.911 & 25.776 & 26.938 & 29.286 \\
                        & $\Delta$ & +3.882 & +2.284 & +2.530 & +3.391 & +4.024 & +4.021 \\
\midrule
\multirow{3}{*}{0.0050} & VR      & 23.168 & 23.277 & 24.513 & 22.625 & 23.004 & 25.482  \\
                        & $\Delta$ &  +3.443 & +3.114 & +2.767 & +3.023 & +3.568 & +3.674  \\
\midrule
\multirow{3}{*}{0.0100} & VR      & 21.032 & 22.360 & 23.161 & 20.966 & 20.721 & 23.223   \\
                        & $\Delta$ & +3.312 & +3.719 & +3.363 & +2.660 & +3.418 & +3.875 \\
\midrule
\multirow{3}{*}{0.0500} & VR      & 14.564 & 17.568 & 19.406 & 14.050 & 16.123 & 18.251 \\
                        & $\Delta$ & +2.916 & +4.527 & +4.380 & +2.086 & +3.724 & +3.993  \\
\midrule
\multirow{3}{*}{0.1000} & VR      & 9.427 & 14.510 & 16.573 & 9.969 & 13.813 & 15.270  \\
                        & $\Delta$ &  +1.376 & +3.538 & +3.725 & +1.842 & +3.478 & +3.353 \\
\bottomrule
\end{tabular}
}
\end{center}
\end{table}

\section{Conclusion}
In this work, we introduced a greedy variable-rate bit allocation algorithm for ACOVQ. This method was applied to a communication system with a Markov noise Polya channel and a source sampled from a one-dimensional Laplacian distribution. Simulation results show that the VR-ACOVQ generated with the proposed bit allocation algorithm significantly outperforms FR-ACOVQ across all tested channel parameters, achieving gains of up to 4.5 dB while never exceeding the average rate of the corresponding FR-ACOVQ schemes. Because the proposed VR-ACOVQ method applies to any channel with memory, similar performance improvements over FR-ACOVQ can be expected for other channel models.

\bibliographystyle{IEEEtran}
\bibliography{sections/ref}

\end{document}